\documentstyle[12pt,epsf]{article}
\begin{document}
\title{NONDEGENERATE FERMIONS
IN THE BACKGROUND OF THE  SPHALERON BARRIER}
\vspace{1.5truecm}
\author{
{\bf Guido Nolte}\\
Fachbereich Physik, Universit\"at Oldenburg, Postfach 2503\\
D-26111 Oldenburg, Germany
\and
{\bf Jutta Kunz}\\
Fachbereich Physik, Universit\"at Oldenburg, Postfach 2503\\
D-26111 Oldenburg, Germany\\
and\\
Instituut voor Theoretische Fysica, Rijksuniversiteit te Utrecht\\
NL-3508 TA Utrecht, The Netherlands
\and
{\bf Burkhard Kleihaus}\\
Fachbereich Physik, Universit\"at Oldenburg, Postfach 2503\\
D-26111 Oldenburg, Germany }
\vspace{1.5truecm}


\maketitle
\vspace{1.0truecm}

\begin{abstract}
We consider level crossing in the background of
the sphaleron barrier for nondegenerate fermions.
The mass splitting within the fermion doublets
allows only for an axially symmetric ansatz for the fermion fields.
In the background of the sphaleron
we solve the partial differential equations for the fermion functions.
We find little angular dependence for our choice of ansatz.
We therefore propose a good approximate
ansatz with radial functions only.
We generalize this approximate ansatz with radial functions only
to fermions in the background of the sphaleron barrier
and argue, that it is a good approximation there, too.
\end{abstract}
\vfill
\vfill\eject

\section{Introduction}

The explanation of the observed baryon asymmetry of the universe
represents a challenging problem.
Although far from solving this highly complex problem,
we know at least what features a theory must have
to allow for an explanation.
It is therefore remarkable
that the standard model fulfills all three
Sakharov-conditions
to generate the observed baryon asymmetry:
C and CP violation, a first order phase transition
and non-conservation of baryon number \cite{sak}.

Here we are concerned with the violation of baryon number
(or more generally fermion number) in the standard model.
It was discovered by 't Hooft \cite{hooft}
as a consequence of the Adler-Bell-Jackiw anomaly
present in chiral gauge theories.
In particular 't Hooft studied the
fermion number violation induced by vacuum to vacuum tunneling
processes described by instantons,
resulting in extremely small tunneling rates.

In Weinberg-Salam theory topologically distinct vacua
are separated by finite energy barriers.
The height of the barriers is given by the energy of the sphaleron,
an unstable solution of the static field equations \cite{man,km}.
Thus the sphaleron determines the minimal energy needed
for a classically allowed vacuum to vacuum transition.
The probability for a transition is expected to be enhanced
significantly, if enough energy is put into the system
under consideration, either in suitable (future) accelerators
or at high temperatures in the early universe [5-10].

While the barrier is traversed baryon number violation may be
seen explicitly by analyzing the corresponding Dirac equation
in the bosonic background fields.
The lowest positive energy continuum state
becomes continuously deformed along the barrier
until it reaches the negative energy continuum,
passing zero energy precisely at the top
of the energy barrier, at the sphaleron [11-15].
Investigating the whole spectrum of the Dirac equation shows,
that along the barrier in fact all levels become
continuously deformed into the next lower levels,
resulting finally in an identical spectrum,
where only the number of occupied
levels above the Dirac sea has decreased by one \cite{dia}.

These calculations [11-16] are based on the approximation,
that the fermion doublets are degenerate in mass (and that
the Weinberg angle may be set to zero \cite{kkb1,kkb2}),
allowing for a spherically symmetric ansatz
for the fermion wave function.
For the physical situation of highly nondegenerate
fermion masses (at least for the heavy flavours)
an analogous calculation is far more involved,
since the spherically symmetric ansatz fails
and the equations of motion cannot be reduced to ordinary
differential equations.
(This is in contrast to the case of instantons \cite{dm1}.)

Here we consider an axially symmetric ansatz
for the fermion fields in the background of the sphaleron barrier.
The ansatz is chosen in such a way,
that it is ``almost spherically symmetric'',
in the sense that the functions involved
have little angular dependence.
Due to the symmetry of the sphaleron
the ansatz simplifies considerably
in the background field of the sphaleron.
In this case we solve numerically the full set of
partial differential equations for the fermion functions.
We then consider a set of approximate ordinary differential equations
for the fermion functions, finding almost identical solutions.
Because of the numerical complexity involved in
solving the full set of partial differential equations
in the background of the sphaleron barrier,
we consider in this general case
only an approximate set of ordinary differential
equations for radial fermion functions.
We argue that these equations represent a good approximation
as well.

In section 2 we briefly review
the Weinberg-Salam Lagrangian (for vanishing
mixing angle) for nondegenerate fermion doublets.
In section 3 we present our axially symmetric ansatz for the fermions,
constructed as a generalization of the usual
spherically symmetric ansatz.
In section 4 we consider fermions in the background of the sphaleron.
We derive the equations of motion, present the solutions
of the full set of partial differential equations,
and compare with the solutions of the set of
approximate ordinary differential equations.
In section 5 we consider
fermions in the background of the sphaleron barrier.
We present our conclusions in section 6.

\section{\bf Weinberg-Salam Lagrangian}

We start with the bosonic sector of the Weinberg-Salam theory
in the limit of vanishing Weinberg angle,
where the electromagnetic field
decouples and can be set to zero,
\begin{equation}
{\cal L}_{\rm b} = -\frac{1}{4} F_{\mu\nu}^a F^{\mu\nu,a}
+ (D_\mu \Phi)^{\dagger} (D^\mu \Phi)
- \lambda (\Phi^{\dagger} \Phi - \frac{1}{2}v^2 )^2
\   \end{equation}
with the field strength tensor
\begin{equation}
F_{\mu\nu}^a=\partial_\mu V_\nu^a-\partial_\nu V_\mu^a
            + g \epsilon^{abc} V_\mu^b V_\nu^c
\ , \end{equation}
and the covariant derivative
\begin{equation}
D_{\mu}  = \partial_{\mu}
             -\frac{1}{2}ig \tau^a V_{\mu}^a
\ . \end{equation}
The ${\rm SU(2)_L}$
gauge symmetry is spontaneously broken
due to the non-vanishing vacuum expectation
value $v$ of the Higgs field
\begin{equation}
    \langle \Phi \rangle = \frac{v}{\sqrt2}
    \left( \begin{array}{c} 0\\1  \end{array} \right)
\ , \end{equation}
leading to the boson masses
\begin{equation}
    M_W = M_Z =\frac{1}{2} g v \ , \ \ \ \ \ \
    M_H = v \sqrt{2 \lambda}
\ . \end{equation}
We employ the values $M_W=80 \ {\rm GeV}$, $g=0.65$.

For vanishing mixing angle,
considering only one fermion doublet,
the fermion Lagrangian reads
\begin{eqnarray}
{\cal L}_{\rm f} & = &
   \bar q_L i \gamma^\mu D_\mu q_L
 + \bar q_R i \gamma^\mu \partial_\mu q_R
   \nonumber \\
           & - & f^{(u)} (\bar q_L
           \tilde \Phi u_R  +\bar u_R \tilde \Phi^\dagger q_L)
               - f^{(d)} (\bar d_R \Phi^\dagger q_L
                         +\bar q_L \Phi d_R)
\ , \end{eqnarray}
where $q_L$ denotes the lefthanded doublet
$(u_L,d_L)$,
while $q_R$ abbreviates the righthanded singlets
$(u_R,d_R)$,
with $\tilde \Phi = i \tau_2 \Phi^*$.
The fermion masses are given by
\begin{equation}
M_{u,d}=\frac{1}{\sqrt{2}}f^{(u,d)} v
\ . \end{equation}

The fermion equations read in dimensionless coordinates
(chosen in units of $M_W$)
\begin{equation}
\big(i\frac{\partial}{\partial t}+i\sigma^i
\frac{\partial}{\partial x^i}+\frac{1}{2}\tau^a V_i^a\sigma^i\big)q_L
-(m M+\Delta m M \tau_z)q_R=0
\label{f1} \end{equation}
and
\begin{equation}
\big(i\frac{\partial}{\partial t}-i\sigma^i
\frac{\partial}{\partial x^i})q_R
-(m M^\dagger+\Delta m \tau_z M^\dagger)q_L=0 \ ,
\label{f2} \end{equation}
where $M$ is the Higgsfield matrix defined by
\begin{equation}
\Phi=\frac{v}{\sqrt{2}}M
\left(\begin{tabular}{c}0\\ 1\end{tabular}\right) \ ,
\end{equation}
and $m$ and $\Delta m$ are the average fermion mass
and half the mass difference (in units of $M_W$)
\begin{equation}
\ m = (M_u+M_d)/(2M_W) \nonumber \ ,
\end{equation}
\begin{equation}
\Delta m = (M_u-M_d)/(2M_W) \nonumber \ .
\end{equation}

\section{Ansatz}
For the gauge and Higgs fields along the sphaleron barrier
we take the usual spherically
symmetric ansatz in the temporal gauge
\begin{eqnarray}
  V_i^a & = & \frac{1-f_A(r)}{gr} \varepsilon_{aij}\hat r_j
  + \frac{f_B(r)}{gr} (\delta_{ia}-\hat r_i \hat r_a)
  + \frac{f_C(r)}{gr} \hat r_i \hat r_a  \ ,
\\
  V_0^a & = & 0\ , \\
    \Phi & = & \frac{v}{\sqrt {2}}
  \Bigl(H(r) + i \vec \tau \cdot \hat r K(r)\Bigr)
    \left( \begin{array}{c} 0\\1  \end{array} \right)
\ . \end{eqnarray}
Due to a residual
gauge degree of freedom we are free to choose the
gauge $f_C=0$.

To construct an appropriate ansatz for nondegenerate fermions
we begin by recalling the
spherically symmetric ansatz for degenerate fermions
with $\Delta m=0$ [11-16,20],
containing four radial functions,
\begin{equation}
q_L(\vec r\,,t) = e^{-i\omega t} M_W^{\frac{3}{2}}
\bigl[ G_L(r)
+ i \vec \sigma \cdot \hat r F_L(r) \bigr] \chi_{\rm h}
\ , \label{ans1} \end{equation}
\begin{equation}
q_R(\vec r\,,t) = e^{-i\omega t} M_W^{\frac{3}{2}}
\bigl[ G_R(r)
+ i \vec \sigma \cdot \hat r F_R(r) \bigr] \chi_{\rm h}
\ , \label{ans2} \end{equation}
where the normalized hedgehog spinor $\chi_{\rm h}$
satisfies the spin-isospin relation
\begin{equation}
\vec \sigma \chi_{\rm h} + \vec \tau \chi_{\rm h} = 0
\ . \end{equation}
The generalized axially symmetric ansatz contains
the spherically symmetric ansatz,
where the four functions $G_L$, $F_L$, $G_R$ and $F_R$
now depend on the variables $r$ and $\theta$.
Because of the presence of the $\tau_z$-terms in the field equations
(\ref{f1})-(\ref{f2}) for $\Delta m \ne 0$,
we need to `double' the ansatz by adding terms of the same structure,
but with $\chi_{\rm h}$ replaced by $\tau_z \chi_{\rm h}$,
involving the four new ($r$ and $\theta$-dependent) functions
$\Delta G_L$, $\Delta F_L$, $\Delta G_R$ and $\Delta F_R$.
The ansatz now contains eight functions,
which are in general complex and $\theta$-dependent,
caused by various occurrences of the nonvanishing anticommutator
$[\vec{\tau}\cdot\hat{r},\tau_z]_+=2\cos{\theta}$
in the equations of motion.
Considering the $\theta$-dependence of the functions,
the real part is even in $\cos\theta$
while the imaginary part is odd.
This then suggests the following parametrization
of the general axially symmetric ansatz,
involving 16 real functions of the
variables $r$ and $p=\cos^2\theta$,
\begin{eqnarray}
q_L(\vec r\,,t)& =& e^{-i\omega t} M_W^{\frac{3}{2}}
\Bigl(
\bigl[ {G_L^1}(r,p)+i\cos(\theta){G_L^2}(r,p)
+ i \vec \sigma \cdot \hat r
({F_L^1}(r,p)+i\cos(\theta){F_L^2}(r,p))
 \bigr]\nonumber\\&&
+\tau_z\bigl[ {\Delta G_L^1}(r,p)+i\cos(\theta){\Delta G_L^2}(r,p)
\nonumber\\&&
+ i \vec \sigma \cdot \hat r
({\Delta F_L^1}(r,p)+i\cos(\theta){\Delta F_L^2}(r,p))
 \bigr]
\Bigr)
\chi_{\rm h}
\ , \label{ans3} \end{eqnarray}
\begin{eqnarray}
q_R(\vec r\,,t)& =& e^{-i\omega t} M_W^{\frac{3}{2}}
\Bigl(
\bigl[ {G_R^1}(r,p)+i\cos(\theta){G_R^2}(r,p)
+ i \vec \sigma \cdot \hat r
({F_R^1}(r,p)+i\cos(\theta){F_R^2}(r,p))
 \bigr]\nonumber\\&&
+\tau_z\bigl[ {\Delta G_R^1}(r,p)+i\cos(\theta){\Delta G_R^2}(r,p)
\nonumber\\&&
+ i \vec \sigma \cdot \hat r
({\Delta F_R^1}(r,p)+i\cos(\theta){\Delta F_R^2}(r,p))
 \bigr]
\Bigr)
\chi_{\rm h} \ .
\  \label{ans4} \end{eqnarray}

The choice of ansatz (\ref{ans3})-(\ref{ans4})
is not unique.
We have also considered alternative parametrizations
of the axially symmetric fermion ansatz.
These involve different fermion functions,
uniquely related to the above fermion functions.
The crucial advantage of the ansatz (\ref{ans3})-(\ref{ans4})
lies in the observation, that its fermion functions
have only a very weak angular dependence
in the background field of the sphaleron, as shown below.
This is in contrast to the alternative parametrizations
considered.

\section{Sphaleron}

We first consider fermions in the background of the sphaleron.
Since the background field barrier is symmetric about the sphaleron,
the fermion eigenvalue is precisely zero at the sphaleron [11-16],
also for nondegenerate fermion masses.
As for degenerate fermion masses,
the fermion ansatz (\ref{ans3})-(\ref{ans4})
then simplifies significantly in the background field
of the sphaleron.
This is due to the parity reflection symmetry of the sphaleron,
for which the functions $f_B$ and $H$ vanish, resulting
in the decoupling of eight of the 16 functions.
These functions, $F_L^1$, $G_L^2$, $\Delta F_L^1$, $\Delta G_L^2$
and $F_R^1$, $G_R^2$, $\Delta F_R^1$, $\Delta G_R^2$,
can therefore consistently be set to zero.
After dropping the number index on the remaining eight functions
the set of partial differential equations in the variables $r$ and $p$
reads
\begin{eqnarray}
0&=&-{G_R}'+\frac{2}{r}p\frac{\partial}{\partial p} {G_R}
+\frac{1}{r}(1+2p\frac{\partial}{\partial p}){\Delta F_R}
-m K{G_L}\nonumber\\&&
+\Delta m K{\Delta G_L}-2pKm{\Delta F_L} \ ,\label{eq1} \\
0&=&-{\Delta G_R}'+\frac{1}{r}(1+2p\frac{\partial}{\partial p}){F_R}
+\frac{2}{r}p\frac{\partial}{\partial p} {\Delta G_R}
+m K{\Delta G_L}\nonumber\\&&
-\Delta m K{G_L}-2pK\Delta m{\Delta F_L} \ ,\label{eq2} \\
0&=&{F_R}'+\frac{1}{r}(3+2p\frac{\partial}{\partial p}){F_R}
+\frac{2}{r}\frac{\partial}{\partial p}{\Delta G_R}
-\Delta m K{\Delta F_L}+m K ({F_L} + 2{\Delta G_L}) \ ,\label{eq3} \\
0&=&{\Delta F_R}'+\frac{1}{r}(3+2p\frac{\partial}{\partial p}){\Delta F_R}
+\frac{2}{r}\frac{\partial}{\partial p}{G_R}
-m K{\Delta F_L}+\Delta m K ({F_L} + 2{\Delta G_L}) \ ,\label{eq4} \\
0&=&{G_L}'-\frac{2}{r}p\frac{\partial}{\partial p} {G_L}
-\frac{1}{r}(1+2p\frac{\partial}{\partial p}){\Delta F_L}
+m K{G_R} +\Delta m K{\Delta G_R} \nonumber\\&&+2pK(m{\Delta F_R}
+\Delta m{F_R})+\frac{1-f_A}{r}({G_L}+p{\Delta F_L}) \ ,\label{eq5} \\
0&=&{\Delta G_L}'
-\frac{2}{r}p\frac{\partial}{\partial p} {\Delta G_L}
-\frac{1}{r}(1+2p\frac{\partial}{\partial p}){F_L}
-K(m {\Delta G_R}+\Delta m {G_R})  \ ,\label{eq6} \\
0&=&-{F_L}'-\frac{1}{r}(3+2p\frac{\partial}{\partial p}){F_L}
-\frac{2}{r}\frac{\partial}{\partial p}{\Delta G_L}
-m K{F_R}-\Delta m K{\Delta F_R} \nonumber\\&&-2K(m{\Delta G_R}
+\Delta m{G_R} )+\frac{1-f_A}{r}({F_L}+{\Delta G_L}) \ ,\label{eq7} \\
0&=&-{\Delta F_L}'-\frac{1}{r}(3+2p\frac{\partial}{\partial p}){\Delta F_L}
-\frac{2}{r}\frac{\partial}{\partial p}{G_L} +\Delta m K{F_R}+m K{\Delta F_R}
\ . \label{eq8} \end{eqnarray}

Inspection of the equations shows, that only three equations,
eqs.~(\ref{eq1}),(\ref{eq2}) and (\ref{eq5}),
contain $p$-dependent terms,
when the terms involving the partial derivative
with respect to $p$, present in all eight equations, are not considered.
In fact only three functions occur with a prefactor $p$.
These are ${F_R}, {\Delta F_R}$ and ${\Delta F_L}$.
If these three functions are small,
then the ansatz is approximately spherically symmetric
in the sense, that all functions have little angular dependence.
In the following we show, that this is indeed the case.

Let us denote the three functions ${F_R}, {\Delta F_R}$ and ${\Delta F_L}$
as $b$, as `bad' functions, and the other five functions
as $g$, as `good' functions.
First we note, that we could set all
three bad functions $b$ consistently equal to zero, if the source term
\begin{equation}
s=-K({F_L}+2{\Delta G_L})
\label{s} \end{equation}
for the bad functions $F_R$ and $\Delta F_R$
in eqs.~(\ref{eq3}) and (\ref{eq4}) did vanish.
Then the five good functions $g$ were pure radial functions.
Let us therefore inspect this source term more closely
and split it into two terms, $s=a_1-a_2$, with
\begin{equation}
\ \  a_1=-K{F_L} \ ,
\end{equation}
and
\begin{equation}
a_2=2K{\Delta G_L} \ .
\end{equation}
If $a_1=a_2$, the source term vanishes.
We now argue that $a_1$ and $a_2$ are approximately equal.
Setting the bad functions ${F_R}, {\Delta F_R}$
and ${\Delta F_L}$ equal to zero,
and neglecting terms with prefactors $\frac{1}{r}$,
for large $r$ eqs.~(\ref{eq6}) and (\ref{eq7})
reduce to
$${\Delta G_L}'=K(m{\Delta G_R}+\Delta m{G_R}) \ , $$
and
$${F_L}'=-2K(m{\Delta G_R}+\Delta m{G_R}) \ . $$
With the proper boundary conditions at infinity
we thus find for large $r$ for the solutions
the desired behaviour, ${F_L}=-2{\Delta G_L}$,
i.e.~the source term vanishes there.
On the other hand, for small $r$ the source term vanishes,
since the function $K$ vanishes.
In the intermediate region the size of the source term
needs numerical analysis.

We have solved the set of partial differential equations
in the background of the sphaleron numerically for various
values of the average mass $m$ and the mass difference $\Delta m$.
Let us consider a typical numerical result.
In Fig.~1 we show the `good' lefthanded functions,
$G_L$, $\Delta G_L$ and $F_L$,
with normalization $G_L(0)=1$,
for three values of the angle $\theta$
($\theta=0$, $\pi/4$ and $\pi/2$)
for the mass parameters $m=0.5$ and $\Delta m=0.25$.
The $\theta$-dependence of the functions is too small to be seen
in the figure, being on the order of $10^{-4}$.
The corresponding bad lefthanded function $\Delta F_L$
is very small, indeed.
For the case considered it is less then $5\cdot 10^{-4}$,
i.e.~two orders of magnitude smaller than the good functions,
with almost no $\theta$-dependence at all.

These results suggest to approximate all functions by radial functions.
We have therefore obtained a new set of ordinary differential equations
by integrating out the $\theta$-dependence in the energy density,
before variation with respect to the fermion functions.
The resulting equations then differ only in prefactors
for the three bad functions, apart from the absence of
the partial derivatives with respect to $p$.
In block form the approximate set of differential equations reads
\begin{equation}
\left(
\begin{tabular}{c}
$g'$ \\
$b'$ \\
\end{tabular}
\right)
=
\left(
\begin{tabular}{cc}
$A$&$B$ \\
$C$&$D$ \\
\end{tabular}
\right)
\left(
\begin{tabular}{c}
$g$ \\
$b$ \\
\end{tabular}
\right)
\ , \end{equation}
where $A$ is a 5 by 5 matrix, $B$ is a 5 by 3 matrix etc.
The vector $Cg$ represents the source terms of the good functions $g$
for the bad functions $b$.
(It is identical in both sets of equations.)
These source terms are $ms$, $\Delta m s$ and zero
for $F_R$, $\Delta F_R$ and $\Delta F_L$, respectively,
with the source $s$ defined in eq.~(\ref{s}).

Solving the approximate set of ordinary differential equations
leads to results almost identical to those
of the full partial differential equations.
This is demonstrated in Fig.~1,
where also the approximate good lefthanded functions
$G_L$, $\Delta G_L$ and $F_L$,
with normalization $G_L(0)=1$, are shown.
The difference of the approximate functions and the exact functions
is too small to be seen in the figure, being on the order of $10^{-3}$.
The bad lefthanded function $\Delta F_L$ is less then $10^{-4}$.

Thus the exact calculation and the radial approximation
result in almost identical results,
and the bad functions are very small, indeed.
We are therefore free to
present in the following only results obtained
with the approximate calculation.
In Figs.~2-4 we show the same good lefthanded functions,
$G_L$, $\Delta G_L$ and $F_L$, as in Fig.~1
for the same value of the average mass $m=0.5$,
but for three different values of the mass difference,
$\Delta m=0.25$, 0.5, and 0.75.
Fig.~5 is the corresponding figure for the good righthanded function $G_R$.
The functions ${G_L}$ and ${G_R}$
are the only functions which do not vanish in the limit $\Delta m=0$.
All other functions, which vanish for
$\Delta m=0$, are approximately proportional to $\Delta m$
as seen in Figs.~3 and 4.
Finally in Fig.~6 we demonstrate the approximate cancellation of the
source terms $a_1$ and $a_2$, responsible for the fact
that the bad functions are very small.

\section{Sphaleron Barrier}

Let us now consider nondegenerate fermions
in the background of the sphaleron barrier.
Along the barrier we expect a smooth transition
of one fermion level from the
positive continuum to the negative continuum.
In the case of degenerate fermion masses,
all fermion levels change along the barrier
to the respective next lower level \cite{dia},
thus only one level crosses zero, and the spectrum
exhibits no crossing of any two levels.
Expecting the same qualitative behaviour
of the spectrum in the case of nondegenerate masses,
the lowest free fermion level, corresponding to the
lower mass fermion of the doublet, then should cross zero.

In the general background of the sphaleron barrier
the full ansatz, eqs.~(\ref{ans3})-(\ref{ans4}), is needed.
The background fields along the barrier
may be taken from the extremal path
calculations \cite{kb3} or, as done here, from
the gradient approach \cite{nk1}.
The set of partial equations for the 16 real fermionic functions
of the variables $r$ and $p$ reads
\begin{eqnarray}
0&=&\omega {F_R^1}-{G_R^1}'+\frac{2}{r}p\frac{\partial}{\partial p} {G_R^1}
+\frac{1}{r}(1+2p\frac{\partial}{\partial p}){\Delta F_R^2}
-m(K{G_L^1}+H{F_L^1})\nonumber\\ &&
-\Delta m(-K{\Delta G_L^1}+H{\Delta F_L^1})-2pKm{\Delta F_L^2}  \ ,
\label{g1} \\
0&=&\omega {F_R^2}-{G_R^2}'+\frac{1}{r}(1+2p\frac{\partial}{\partial p}){G_R^2}
-\frac{2}{r}\frac{\partial}{\partial p}{\Delta F_R^1}
-\Delta m(-K{\Delta G_L^2}+H{\Delta F_L^2}) \nonumber\\ &&
-m(K{G_L^2}+H{F_L^2}-2K{\Delta F_L^1})  \ , \label{g2} \\
0&=&\omega {\Delta F_R^1}-{\Delta G_R^1}'+\frac{1}{r}(1+2p\frac{\partial}
{\partial p}){F_R^2}+\frac{2}{r}p\frac{\partial}{\partial p}
{\Delta G_R^1}- m(-K{\Delta G_L^1}+H{\Delta F_L^1})\nonumber\\ &&
-\Delta m(K{G_L^1}+H{F_L^1})-2pK\Delta m{\Delta F_L^2} \ , \label{g3} \\
0&=&\omega {\Delta F_R^2}-{\Delta G_R^2}'+\frac{1}{r}(1+2p\frac{\partial}
{\partial p}){\Delta G_R^2}-\frac{2}{r}\frac{\partial}{\partial p}
{F_R^1}- m(-K{\Delta G_L^2}+H{\Delta F_L^2})\nonumber\\ &&
-\Delta m(K{G_L^2}+H{F_L^2}-2K{\Delta F_L^1}) \ , \label{g4} \\
0&=&\omega {G_R^1} +{F_R^1}'
+\frac{1}{r}(2+2p\frac{\partial}{\partial p}){F_R^1}
-\frac{1}{r}(1+2p\frac{\partial}{\partial p}){\Delta G_R^2}
-m(H{G_L^1}-K{F_L^1})\nonumber\\ &&
-\Delta m(H{\Delta G_L^1}+K{\Delta F_L^1})-2pKm{\Delta G_L^2} \ , \label{g5} \\
0&=&\omega {G_R^2}+{F_R^2}'+\frac{1}{r}(3+2p\frac{\partial}{\partial p}){F_R^2}
+\frac{2}{r}\frac{\partial}{\partial p}{\Delta G_R^1}
-\Delta m(H{\Delta G_L^2}+K{\Delta F_L^2}) \nonumber\\ &&
-m(H{G_L^2}-K{F_L^2} -2K{\Delta G_L^1})\ , \label{g6} \\
0&=&\omega {\Delta G_R^1}+{\Delta F_R^1}'
+\frac{1}{r}(2+2p\frac{\partial}{\partial p}){\Delta F_R^1}
-\frac{1}{r}(1+2p\frac{\partial}{\partial p}){G_R^2}
-\Delta m(H{G_L^1}-K{F_L^1})\nonumber\\ &&
-m(H{\Delta G_L^1}+K{\Delta F_L^1})-2pK\Delta m{\Delta G_L^2} \ , \label{g7} \\
0&=&\omega {\Delta G_R^2}+{\Delta F_R^2}'
+\frac{1}{r}(3+2p\frac{\partial}{\partial p}){\Delta F_R^2}
+\frac{2}{r}\frac{\partial}{\partial p}{G_R^1}
-m(H{\Delta G_L^2}+K{\Delta F_L^2}) \nonumber\\ &&
-\Delta m(H{G_L^2}-K{F_L^2}-2K{\Delta G_L^1}) \ , \label{g8} \\
0&=&\omega {F_L^1}+{G_L^1} '-\frac{2}{r}p\frac{\partial}{\partial p} {G_L^1}
-\frac{1}{r}(1+2p\frac{\partial}{\partial p}){\Delta F_L^2}
+m(K{G_R^1} -H{F_R^1})
+\Delta m(K{\Delta G_R^1}-H{\Delta F_R^1})\nonumber\\ &&
+2pK(m{\Delta F_R^2}
+\Delta m{F_R^2})+\frac{1-f_A}{r}({G_L^1}+p{\Delta F_L^2})+\frac{f_B}{r}(
{F_L^1}-p{\Delta G_L^2}) \ , \label{g9} \\
0&=&\omega {F_L^2}+{G_L^2}'
-\frac{1}{r}(1+2p\frac{\partial}{\partial p}){G_L^2}+
\frac{2}{r}\frac{\partial}{\partial p} {\Delta F_L^1}
+ m(K{G_R^2}-H{F_R^2})
+\Delta m(K{\Delta G_R^2}-H{\Delta F_R^2})\nonumber\\ &&
-2K(m{\Delta F_R^1}
+\Delta m{F_R^1})+\frac{1-f_A}{r}({G_L^2}-{\Delta F_L^1})+\frac{f_B}{r}(
{F_L^2}+{\Delta G_L^1}) \ , \label{g10} \\
0&=&\omega {\Delta F_L^1}+{\Delta G_L^1}'
-\frac{1}{r}(1+2p\frac{\partial}{\partial p})
{F_L^2}-\frac{2}{r}p\frac{\partial}{\partial p} {\Delta G_L^1}-
 m(K{\Delta G_R^1}+H{\Delta F_R^1})\nonumber\\ &&
-\Delta m(K{G_R^1} +H{F_R^1}) \ , \label{g11} \\
0&=&\omega {\Delta F_L^2}+{\Delta G_L^2}'
-\frac{1}{r}(1+2p\frac{\partial}{\partial p})
{\Delta G_L^2}+\frac{2}{r}\frac{\partial}{\partial p} {F_L^1} \nonumber\\&&
-m(K{\Delta G_R^2} +H{\Delta F_R^2})-\Delta m(K{G_R^2}+H{F_R^2}) \ ,
\label{g12} \\
0&=&\omega {G_L^1}-{F_L^1}'-\frac{1}{r}(2+2p\frac{\partial}{\partial p}){F_L^1}
+\frac{1}{r}(1+2p\frac{\partial}{\partial p}){\Delta G_L^2}
 -m(H{G_R^1} +K{F_R^1})
-\Delta m(H{\Delta G_R^1}\nonumber\\ &&
+K{\Delta F_R^1})
+2pK(m{\Delta G_R^2}
+\Delta m{G_R^2})+\frac{1-f_A}{r}({F_L^1}-p{\Delta G_L^2})
-\frac{f_B}{r}({G_L^1}+p{\Delta F_L^2}) \ , \label{g13} \\
0&=&\omega {G_L^2}-{F_L^2}'-\frac{1}{r}(3+2p\frac{\partial}{\partial p}){F_L^2}
-\frac{2}{r}\frac{\partial}{\partial p}{\Delta G_L^1}
 -m(H{G_R^2}+K{F_R^2})
-\Delta m(H{\Delta G_R^2}+K{\Delta F_R^2})\nonumber\\ &&
-2K(m{\Delta G_R^1} +\Delta m{G_R^1} )
+\frac{1-f_A}{r}({F_L^2}+{\Delta G_L^1})
+\frac{f_B}{r}( -{G_L^2}+{\Delta F_L^1}) \ , \label{g14} \\
0&=&\omega {\Delta G_L^1}-{\Delta F_L^1}'
-\frac{1}{r}(2+2p\frac{\partial}{\partial p}){\Delta F_L^1}
+\frac{1}{r}(1+2p\frac{\partial}{\partial p}){G_L^2}
-\Delta m(H{G_R^1} -K{F_R^1})\nonumber\\ &&
-m(H{\Delta G_R^1}-K{\Delta F_R^1})  \ , \label{g15} \\
0&=&\omega {\Delta G_L^2}-{\Delta F_L^2}'
-\frac{1}{r}(3+2p\frac{\partial}{\partial p})
{\Delta F_L^2} -\frac{2}{r}\frac{\partial}{\partial p}{G_L^1}
-\Delta m(H{G_R^2}-K{F_R^2})\nonumber\\ &&
-m(H{\Delta G_R^2}-K{\Delta F_R^2}) \ . \label{g16}
\end{eqnarray}

These equations are analogous in structure to the equations in the
sphaleron background, with all relevant features `doubled'.
Now six equations contain $p$-dependent terms (apart from the
terms containing partial derivatives with respect to $p$).
These are eqs.~(\ref{g1}), (\ref{g3}), (\ref{g5}), (\ref{g7}),
(\ref{g9}) and (\ref{g13}).
And six functions occur with a prefactor $p$,
these are ${F_R^2}, {\Delta F_R^2}$, ${\Delta F_L^2}$,
${G_R^2}, {\Delta G_R^2}$ and ${\Delta G_L^2}$,
the six `bad' functions, $b$.
The other ten functions are the `good' functions, $g$.
Again, if the bad functions are small,
all functions have little angular dependence,
and an approximation with radial functions only
will be good.

Let us therefore inspect the two source terms for the bad functions,
\begin{equation}
s_1= H{G_L^2}-K{F_L^2}-2K{\Delta G_L^1} \ ,
\end{equation}
\begin{equation}
s_2= K{G_L^2}+H{F_L^2}-2K{\Delta F_L^1} \ ,
\end{equation}
occurring in eqs.~(\ref{g6}), (\ref{g8}), and in (\ref{g2}), (\ref{g4}),
respectively,
and split these two source terms according to
$s_1=a_1-a_2$, with
\begin{equation}
\ \ \ a_1=H{G_L^2}-K{F_L^2} \ ,
\end{equation}
\begin{equation}
a_2=2K{\Delta G_L^1} \ ,
\end{equation}
and $s_2=b_1-b_2$, with
\begin{equation}
\ \ b_1=K{G_L^2}+H{F_L^2} \ ,
\end{equation}
\begin{equation}
b_2=2K{\Delta F_L^1} \ .
\end{equation}
If both source terms are small,
then the bad functions are small,
and consequently the angular dependence of all 16
fermion functions is small.

Due to its great complexity,
we have not yet attempted to solve the full set of 16 coupled
partial differential equations numerically.
Instead we have from the beginning resorted
to the study of the approximate set of 16 ordinary
differential equations, obtained by integrating out the
angular dependence in the energy density.
But even this approximate set of 16 ordinary
differential equations has resisted a numerical solution
along the full sphaleron barrier.
Only by setting two of the 16 radial functions
explicitly to zero,
namely the supposedly small bad functions
${\Delta G_L^2}$ and ${\Delta G_R^2}$,
we have succeeded in constructing the fermion solution
along the sphaleron barrier.
(Note, that $\Delta G_L^2$ has no source term.)

Without the solution of the partial diffential equations
to compare with,
the quality of the approximate solution is not known
along the full barrier, away from the sphaleron.
At the sphaleron the approximation is excellent,
and it should remain good close to the sphaleron.
Away from the sphaleron, however, we can at least make a
consistency check for the radial approximation used,
by inspecting the source terms $s_1$ and $s_2$
in this approximation.
Numerical analysis shows, that the source terms are indeed small.
In Fig.~7 we show as a typical example along the barrier
the source terms $b_1$ and $b_2$
for the Chern-Simons number $N_{CS}=0.4$
and the mass parameters $m=0.5$ and $\Delta m=0.25$.
While the cancellation of the source terms $a_1$ and $a_2$
remains as good along the barrier as it is at the sphaleron
(shown in Fig.~6),
the cancellation of the additional source terms $b_1$ and $b_2$
is even much better.
This indicates, that the bad functions are indeed small compared to
the good functions.
The radial approximation therefore should be good along the full
sphaleron barrier.

Let us then discuss the level crossing along the sphaleron
barrier, as obtained with the approximate radial set of equations.
In Fig.~8 we present the fermion eigenvalue along the barrier for
an average mass of $m=2$ and for several values
of the mass difference, $\Delta m =0.5$, 1.0 and 1.5.
The eigenvalue starts from the positive continuum at the lower mass
(1.5, 1.0 and 0.5, respectively), and reaches the negative continuum
at the corresponding negative value.
The bigger the mass splitting, i.e.~the smaller the lower mass,
the later the fermion level leaves the continuum
to become bound, analogous to the case of
degenerate fermion masses [14,15,20].

For degenerate fermion masses the fermion wavefunction
is determined by the hedgehog spinor $\chi_{\rm h}$,
giving both isospin components of the fermion doublet
an equal amplitude along the sphaleron barrier.
For nondegenerate fermion masses this is no longer the case.
Let us define the up-part of the fermion wavefunction
along the barrier as
\begin{equation}
\frac{<P\Psi,P\Psi>}{<\Psi,\Psi>} \ ,
\end{equation}
where $P$ projects out the upper isospin component.
(Note, that this definition of the up-part is not gauge invariant.)
For degenerate fermions the up-part is everywhere one half.
For nondegenerate fermions the up-part along the barrier depends
on the size of the mass splitting,
as shown in Fig.~9 (for the mass parameters employed also in Fig.~8).
The up-part dominates slightly in the vicinity of the sphaleron
and clearly disappears when the vacua are reached.
Remarkably, the point where the down-part equals the up-part
only depends on $m$ and not on $\Delta m$.

We finally address the question,
how to best approximate a fermion solution
in the physical situation of nondegenerate fermion masses
by a far simpler solution, obtained in the approximation
of degenerate fermion masses,
in the vicinity of the sphaleron.
In the physical case
the mass parameters $m$ and $\Delta m$ determine the
nondegenerate masses, $m+\Delta m$ and $m-\Delta m$.
Close to the continuum clearly the lower mass, $m-\Delta m$,
is the relevant fermion mass.
In the vicinity of the sphaleron, however,
it is the average mass $m$, which matters.
In fact, the average mass $m$ of the nondegenerate case
mostly leads to an excellent approximation
for the fermion eigenvalue in the vicinity of the sphaleron,
when employed in the far simpler calculations
with degenerate fermion mass.
This is demonstrated in Fig.~10,
where we compare the nondegenerate case $m=2$, $\Delta m=1$
with the degenerate cases $m=1$, $\Delta m=0$ and
$m=2$, $\Delta m=0$.
Having the same average mass,
the fermion eigenvalues in the nondegenerate case,
and in the second degenerate case,
agree very well in the vicinity of the sphaleron.
In Fig.~11 we present the slope of the fermion eigenvalue
at the sphaleron as a function of the mass difference,
for three values of the average mass, $m=0.5$, 1 and 2.
We observe, that the slope is
fairly independent of the mass difference $\Delta m$
for not too large values of the average mass $m$.

\section{Conclusion}

We have considered level crossing in the background field of the sphaleron
barrier for fermion doublets with nondegenerate masses.
The mass splitting necessitates a generalized ansatz
for the fermions, possessing only axial symmetry.
We have proposed a particular parametrization of the
axially symmetric ansatz,
containing 16 real functions of the two variables $r$ and $p=\cos^2\theta$.
The structure of the ansatz chosen is based on the
structure of the spherically symmetric ansatz,
which represents its simple limit
for vanishing mass splitting.
This particular parametrization has the great advantage,
that it leads to fermion functions
with little angular dependence in the background
of the sphaleron,
and (supposedly) also along the full sphaleron barrier.

In the background field of the sphaleron
the proposed ansatz simplifies considerably.
It leads to a set of eight partial differential equations.
We have solved these equations numerically,
finding that the resulting fermion functions
have very little angular dependence.
The reason lies in the structure of the equations
for this particular choice of ansatz.
Only three functions occur with an angular dependent
prefactor $p$ (apart from partial derivative terms),
and there is a single source term for these three functions.
Since this source term is small, these three functions,
which introduce explicit angular dependence into the equations,
are small, and consequently all eight functions have only little
angular dependence.

We have then proposed an approximate ansatz
with radial functions only.
Integrating out the angular dependence in the energy density,
leads to a new approximate set of ordinary differential
equations. Solving these numerically,
we find that the solutions are in excellent agreement with those
of the full calculation.
Thus we have an excellent radial approximation
for nondegenerate fermion masses at the sphaleron.

In the general case of fermions in the background of the sphaleron
barrier, we have found the same structure of the
equations as in the sphaleron case, but with all relevant
features `doubled', since the ansatz no longer simplifies.
As yet we have only solved the
approximate set of ordinary differential equations,
obtained by integrating out the angular dependence
in the energy density (and then setting two of the small functions
explicitly to zero).
Without the solution of the set of partial differential
equations to compare with, we do not know the quality of
the approximation away from the sphaleron.
However, we have made a consistency check
by evaluating the two source terms for the six functions,
which introduce explicit angular dependence into the equations.
Since the source terms are small, these functions are small,
and consequently all functions have only little angular dependence.
We therefore argue, that the radial approximation employed
should be good along the full barrier.

Considering level crossing along the barrier,
we have observed that the fermion mode which crosses zero energy
at the sphaleron reaches the continua at the lower
fermion mass, as expected.
Finally we have shown, that in the vicinity of the sphaleron the
eigenvalue for nondegenerate fermions with average mass $m$
and mass difference $\Delta m$,
may be well approximated by
the eigenvalue obtained with the far simpler calculation,
involving only degenerate fermions with the average mass $m$.
With respect to the large splitting
of the top and bottom quark masses, this suggests
to rather use half the top quark mass in approximate
calculations with degenerate fermions.

\newpage


\begin{figure}
\epsfxsize=8cm
\epsffile[-100 -0 250 300]{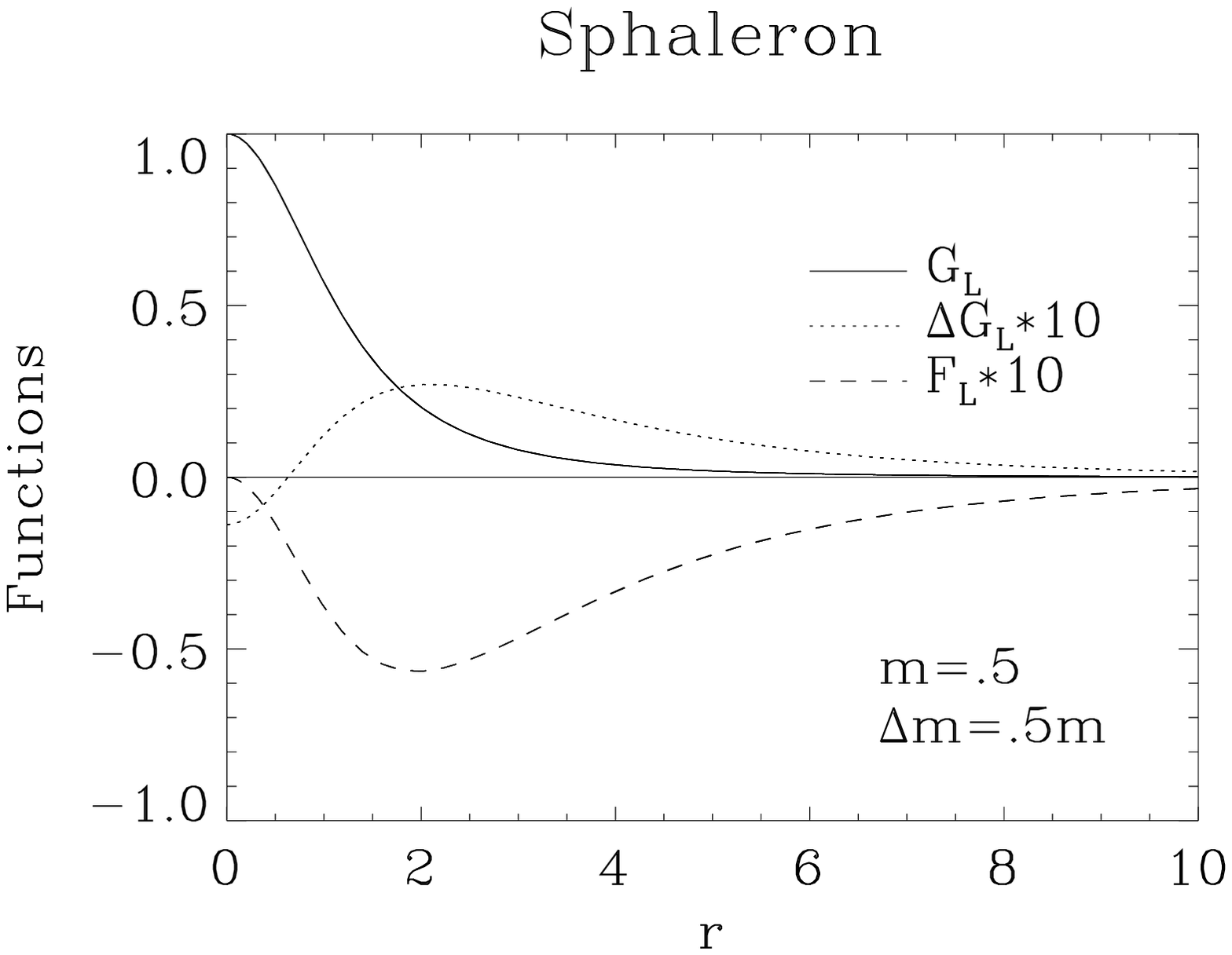}
\caption{
The `good' lefthanded functions,
$G_L$ (solid), $\Delta G_L$ (dotted) and $F_L$ (dashed),
in the background field of the sphaleron
with normalization $G_L(0)=1$,
in the exact calculation
for three values of the angle $\theta$
($\theta=0$, $\pi/4$ and $\pi/2$)
and in the approximate calculation,
with the mass parameters $m=0.5$ and $\Delta m=0.25$.
Any of the three visible lines consists of four individual lines.
}
\end{figure}

\begin{figure}
\epsfxsize=8cm
\epsffile[-100 -0 250 300]{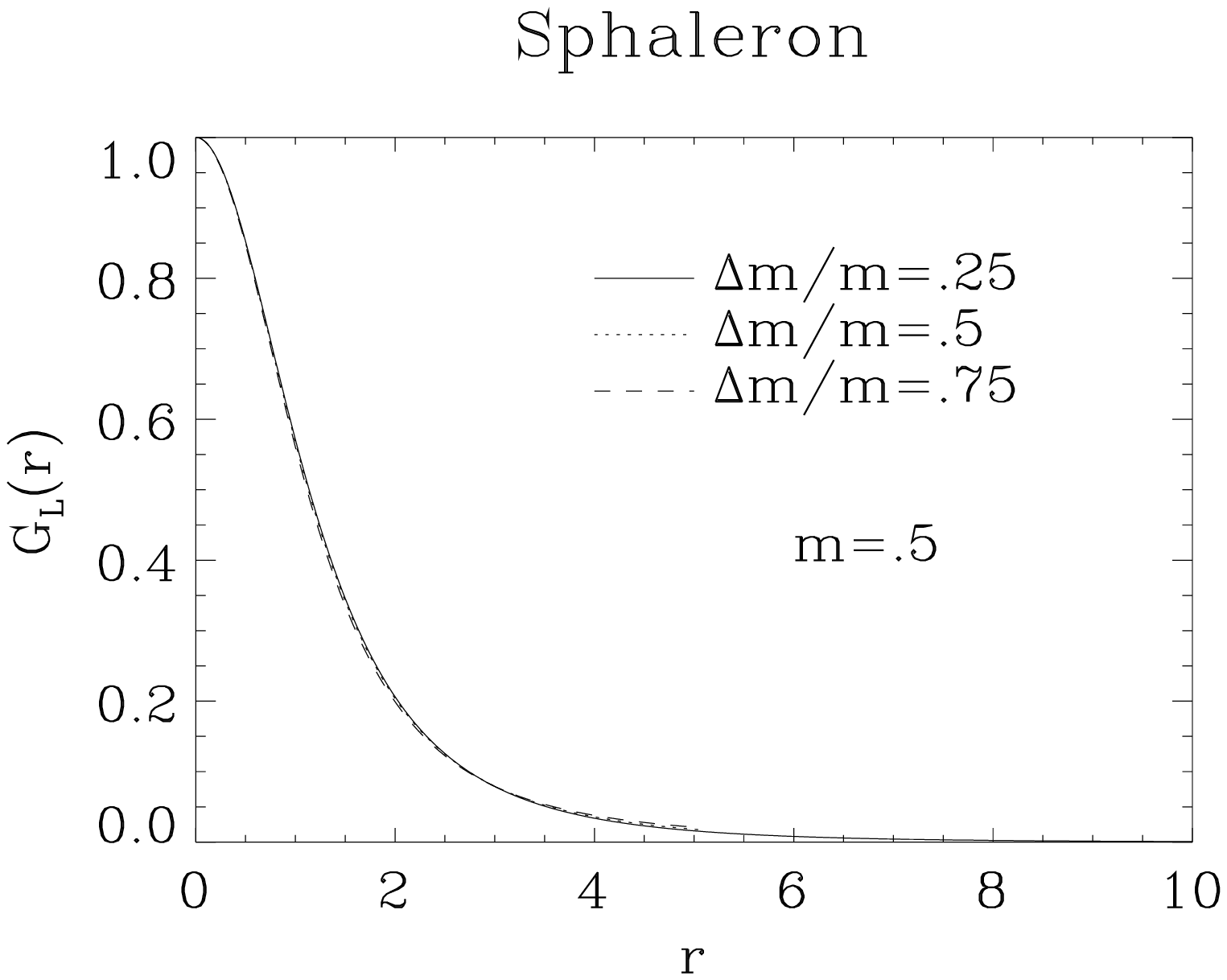}
\caption{
The `good' lefthanded function $G_L$
in the background field of the sphaleron
in the approximate calculation,
for the fixed average mass $m=0.5$ and
three values of the mass difference
$\Delta m=0.25$ (solid),
$\Delta m=0.50$ (dotted),
$\Delta m=0.75$ (dashed).
}
\end{figure}

\begin{figure}
\epsfxsize=8cm
\epsffile[-100 -0 250 300]{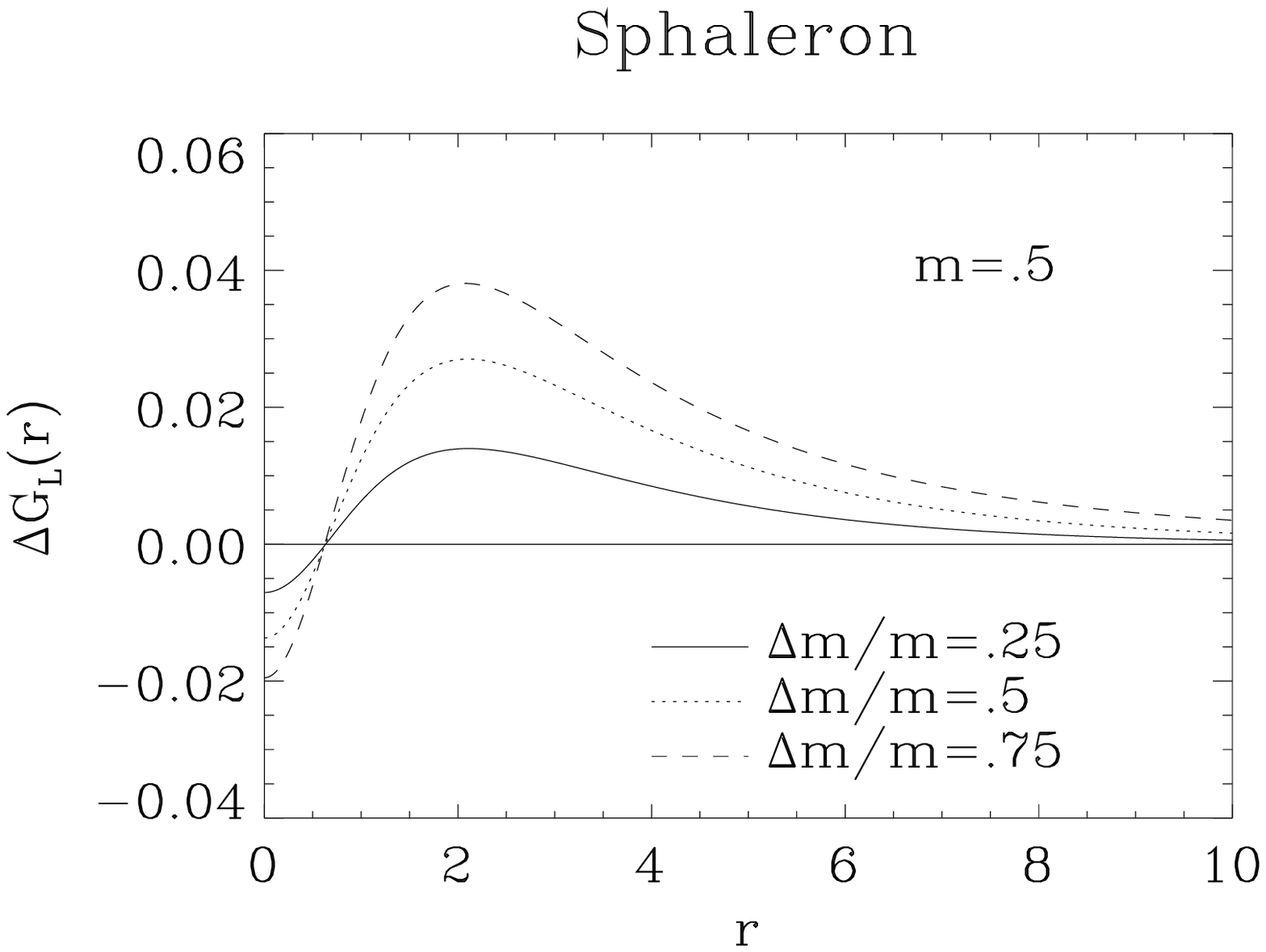}
\caption{
Same as Fig.~2 for ${\Delta G_L}$.
}
\end{figure}

\begin{figure}
\epsfxsize=8cm
\epsffile[-100 -0 250 300]{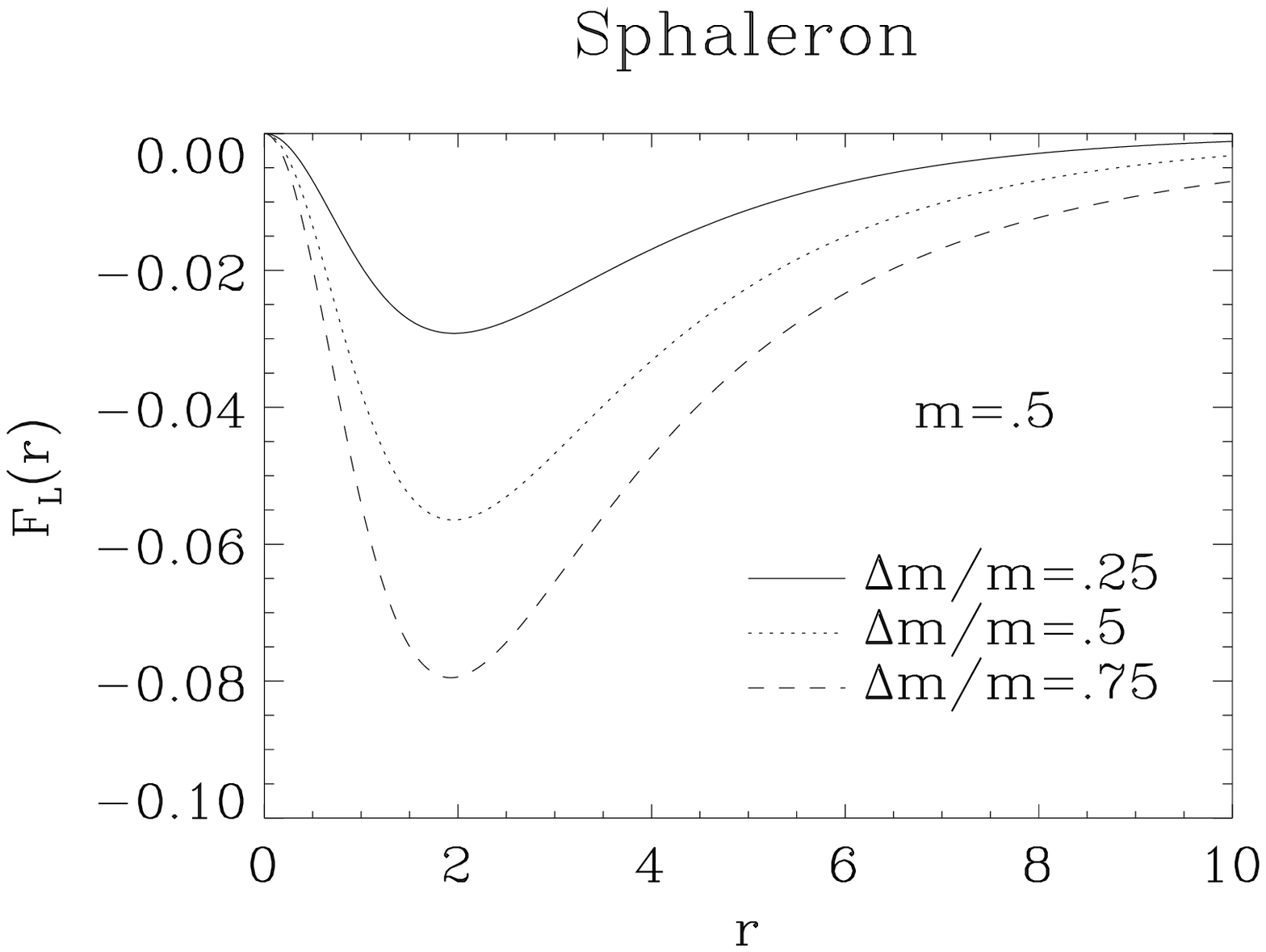}
\caption{
Same as Fig.~2 for ${F_L}$.
}
\end{figure}\newpage

\begin{figure}
\epsfxsize=8cm
\epsffile[-100 -0 250 300]{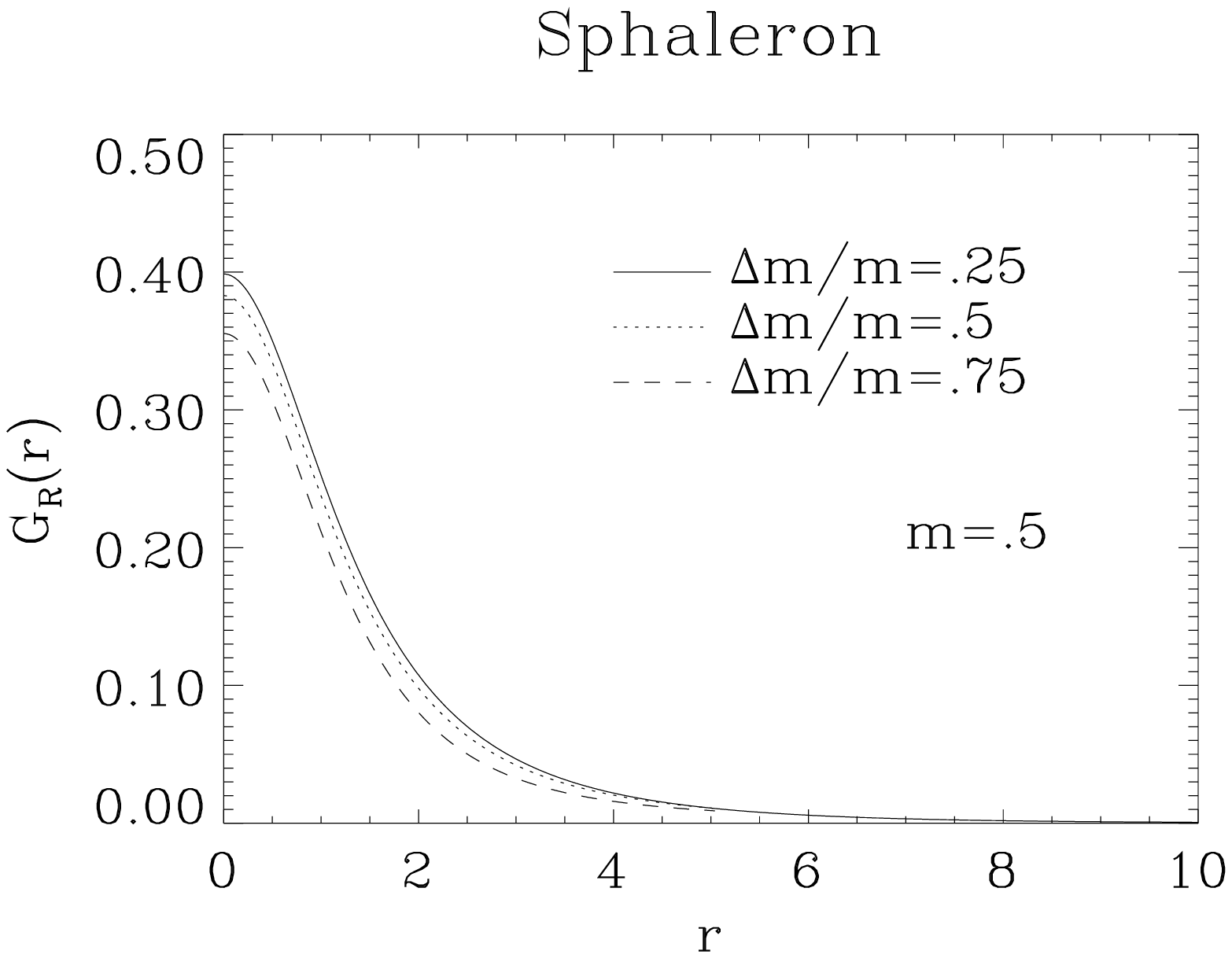}
\caption{
Same as Fig.~2 for ${G_R}$.
}
\end{figure}

\begin{figure}
\epsfxsize=8cm
\epsffile[-100 -0 250 300]{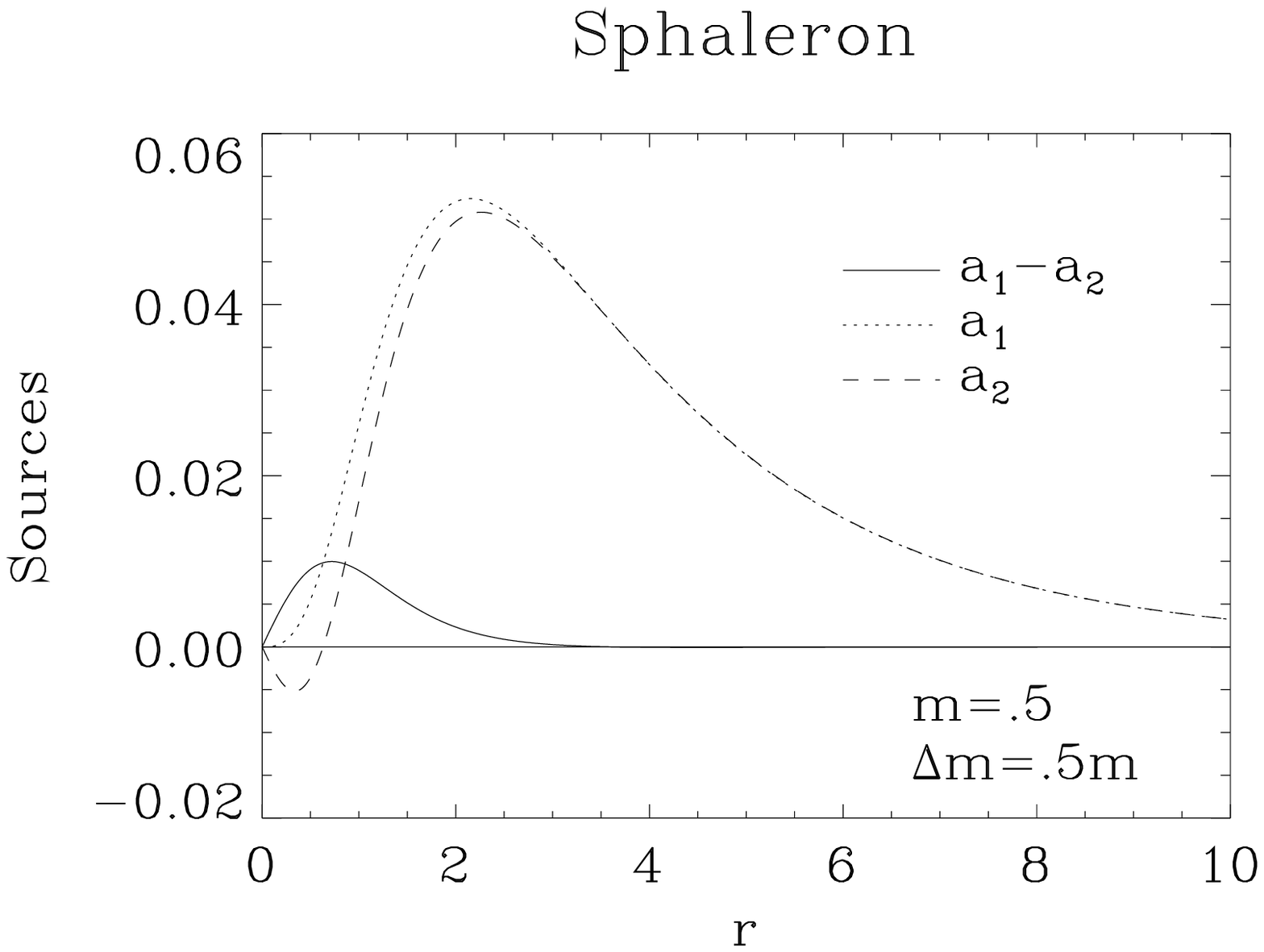}
\caption{
The source term $s=a_1-a_2$ (solid), and its individual parts
$a_1$ (dotted) and $a_2$ (dashed)
in the background field of the sphaleron
in the approximate calculation,
with the mass parameters $m=0.5$ and $\Delta m=0.25$.
}
\end{figure}

\begin{figure}
\epsfxsize=8cm
\epsffile[-100 -0 250 300]{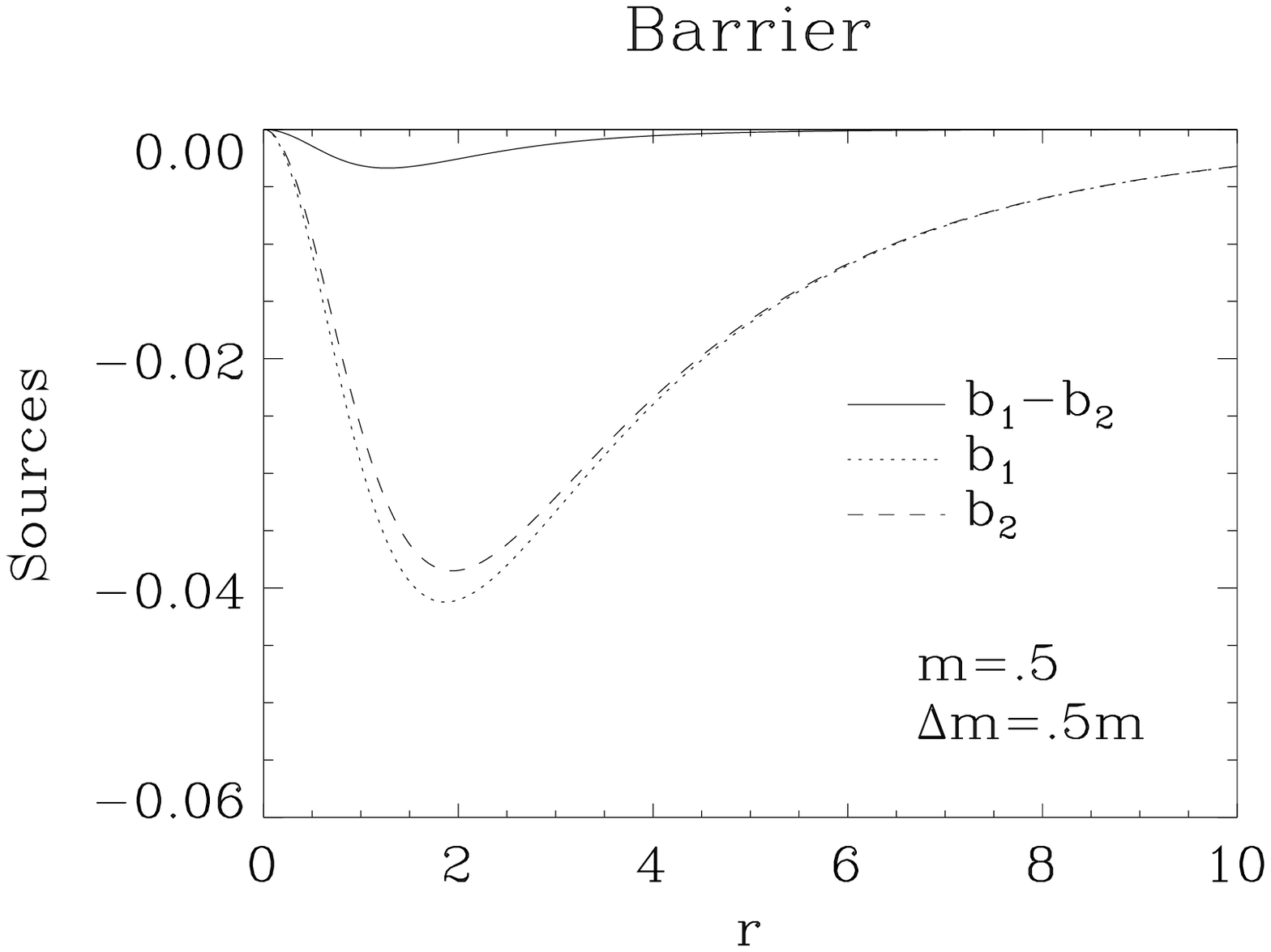}
\caption{
The source term $s_2=b_1-b_2$ (solid), and its individual parts
$b_1$ (dotted) and $b_2$ (dashed)
in the background field of the sphaleron barrier
at the Chern-Simons number $N_{CS}=0.4$
in the approximate calculation,
with the mass parameters $m=0.5$ and $\Delta m=0.25$.
}
\end{figure}\newpage

\begin{figure}
\epsfxsize=8cm
\epsffile[-100 -0 250 300]{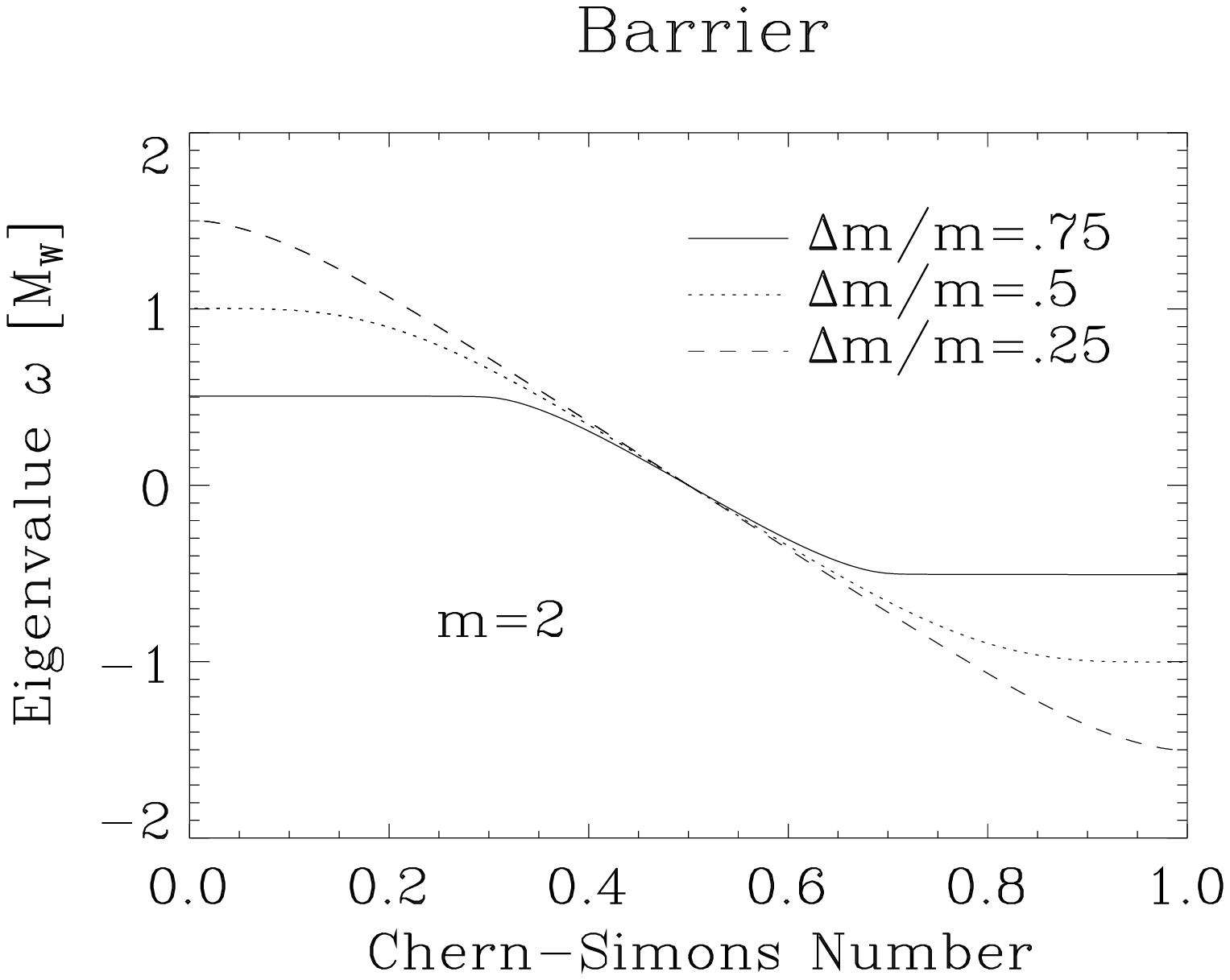}
\caption{
The fermion eigenvalue along the sphaleron barrier
in the approximate calculation,
for the fixed average mass $m=2$ and
three values of the mass difference
$\Delta m=0.75$ (solid),
$\Delta m=0.50$ (dotted),
$\Delta m=0.25$ (dashed).
}
\end{figure}

\begin{figure}
\epsfxsize=8cm
\epsffile[-100 -0 250 300]{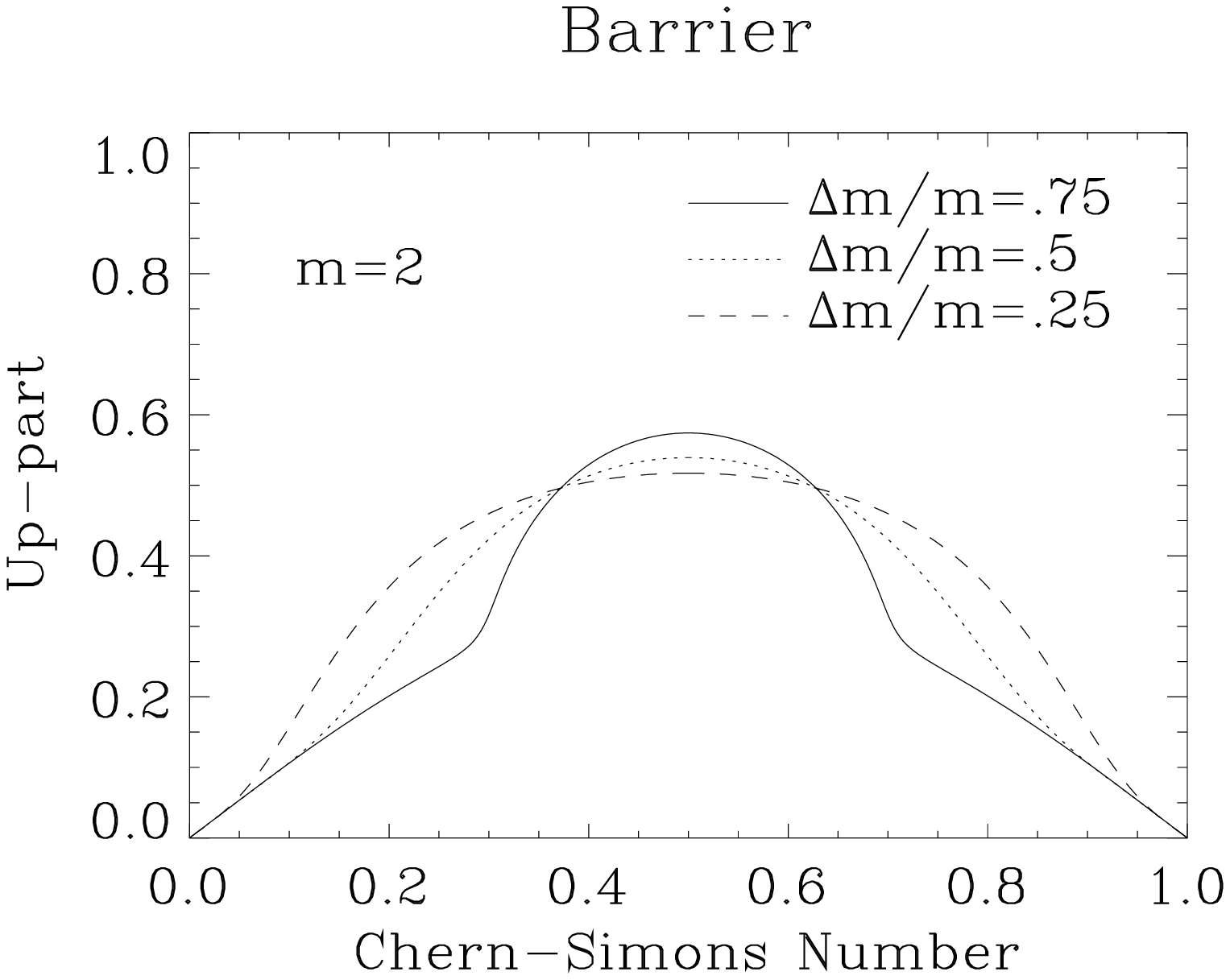}
\caption{
The up-part of the fermion wavefunction along the sphaleron barrier
in the approximate calculation,
for the fixed average mass $m=2$ and
three values of the mass difference
$\Delta m=0.75$ (solid),
$\Delta m=0.50$ (dotted),
$\Delta m=0.25$ (dashed).
}
\end{figure}

\begin{figure}
\epsfxsize=8cm
\epsffile[-100 -0 250 300]{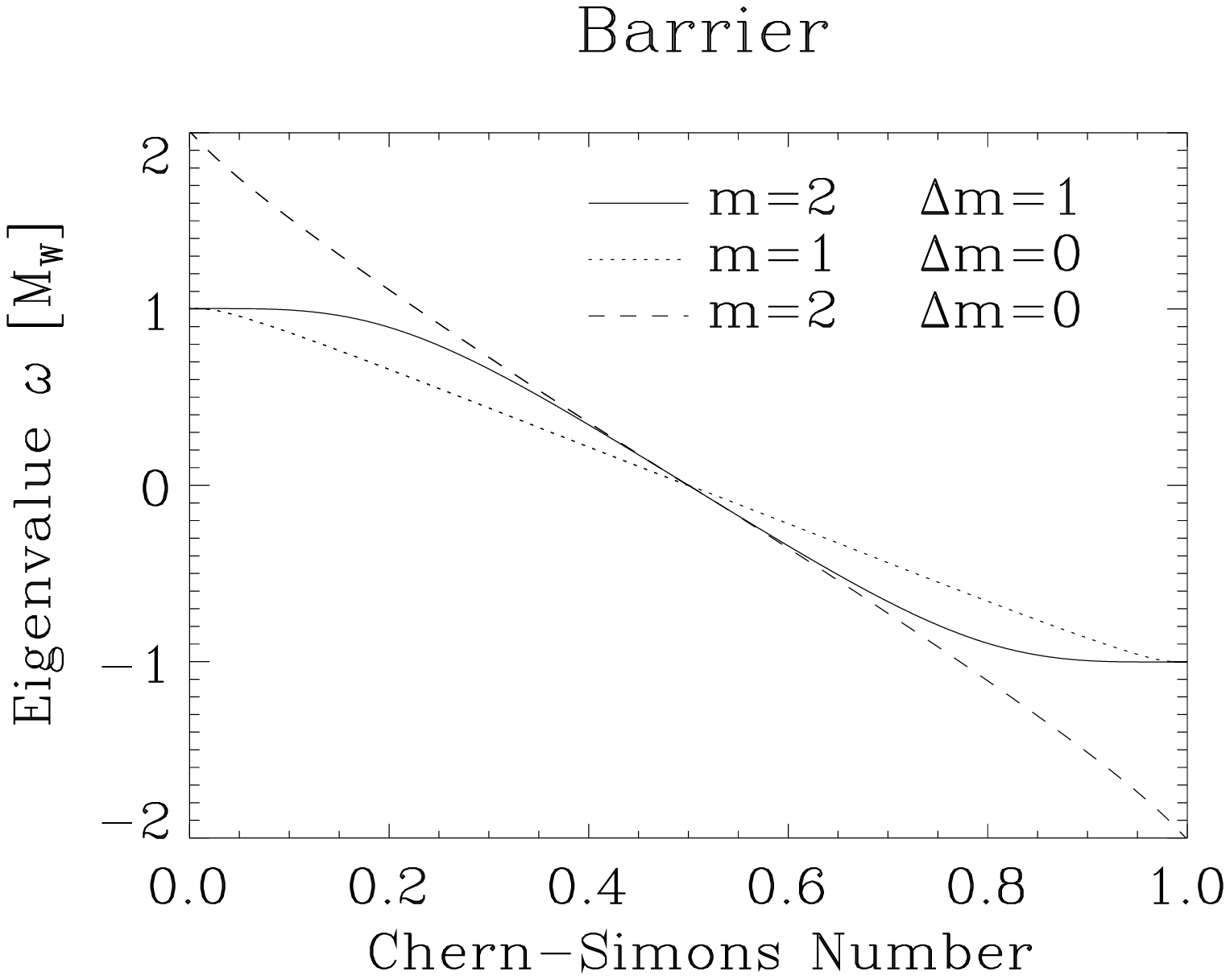}
\caption{
The fermion eigenvalue along the sphaleron barrier
in the approximate calculation
for the average mass
$m=2$ and the mass difference $\Delta m=1$ (solid),
compared to the fermion eigenvalue
for degenerate fermion masses for
$m=1$ and $\Delta m=0$ (dotted), and
$m=2$ and $\Delta m=0$ (dashed).
}
\end{figure}

\begin{figure}
\epsfxsize=8cm
\epsffile[-100 -0 250 300]{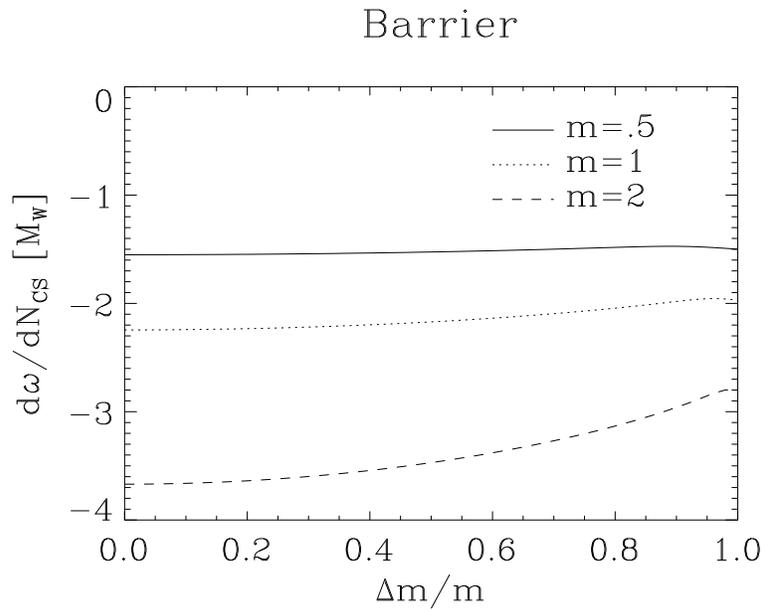}
\caption{
The slope of the fermion eigenvalue at the sphaleron
in the approximate calculation,
as a function of the mass difference $\Delta m$
for three values of the average mass,
$m=0.5$ (solid),
$m=1$ (dotted),
$m=2$ (dashed).
}
\end{figure}

\end{document}